\documentclass[aps,preprint]{revtex4}%
\usepackage{amsfonts}
\usepackage{amsmath}
\usepackage{amssymb}
\usepackage{graphicx}%
\providecommand{\U}[1]{\protect\rule{.1in}{.1in}}

\begin{document}
\preprint{ }
\title[QED$_{3}$]{Infrared behaviour of the fermion propagator in unquenched QED$_{3}$ with
finite threshold effects}
\author{Yuichi Hoshino}
\affiliation{Kushiro National College of Technology,Otanoshike Nishi 2-32-1,Kushiro
City,HOkkaido 084,Japan}
\pacs{PACS number}

\begin{abstract}
To remove the linear infrared divergences in quenched approximation we include
the massive fermion loop to the photon spectral function.Spectral function of
fermion has no one particle singularity if we fix the anomalous dimension to
be unity.In the case of $N$ flavour,$N$ dependence of order parameter is mild
which may be due to screening effects.

\end{abstract}
\eid{identifier}
\date{}
\startpage{101}
\endpage{102}
\maketitle
\tableofcontents

\section{Introduction}

To study the infrared behavior of the propagator in QED$_{2+1}$,we evaluated
the spectral function which is known as the Bloch-Nordsieck approximation in
four dimension.In quenched case linear and logarithmic infrared divergences
appear in the order $e^{2}$ spectral function $F$.Exponentiation of $F$ yields
full propagator with all order of infrared divergences.In four dimension
anomalous dimension modifies the short distance singularity at least for weak
coupling,which leads cut structure near the mass shell.For fixed infrared
cut-off we expand the function $F$ and find a mass shift,its log correction
and anomalous dimension of wave function.To avoid the infrared divergences we
choose the gauge $d=-1$ and include effects of massive fermion loops to the
photon spectral function.The resutls have no infrared divergence but the
effects of finite threshold seems to make oscillation of the propagator.If we
set the anomalous dimension to be unity short distance singularity
disappears,chiral symmetry breaks dynamically, and the spectral function has
$-\delta^{\prime}(p^{2}/m^{2}-1)$ like singularity which shows the absence of
one particle state and vanishment of $Z_{2}^{-1}$.

\section{Spectral representation of the propagator}

\subsection{\bigskip Fermion}

In this section we show how to evalute the fermion propagator non
pertubatively by the spectral represntation which preserves unitarity and
analyticity [1,2,3]. Assuming parity conservation we adopt 4-component spinor.
The spectral function of the fermion in (2+1) dimension is defined
\begin{align}
\left\langle 0|T(\psi(x)\overline{\psi}(y)|0\right\rangle  &  =i\int
\frac{d^{3}p}{(2\pi)^{3}}e^{-ip\cdot(x-y)}\int_{0}^{\infty}ds\frac{\gamma\cdot
p\rho_{1}(s)+\rho_{2}(s)}{p^{2}-s+i\epsilon},\\
\rho(p)  &  =-\frac{1}{\pi}\operatorname{Im}S_{F}(p)=\gamma\cdot p\rho
_{1}(p)+\rho_{2}(p)\nonumber\\
&  =(2\pi)^{2}\sum_{n}\delta^{(3)}(p-p_{n})\left\langle 0|\psi
(x)|n\right\rangle \left\langle n|\overline{\psi}(0)|0\right\rangle .
\end{align}
In the quenched approximation the state $|n>$ stands for a fermion and
arbitrary numbers of photons,%
\begin{equation}
|n>=|r;k_{1},...,k_{n}>,r^{2}=m^{2}.
\end{equation}
In deriving the matrix element $\left\langle 0|\psi(x)|n\right\rangle $ we
must take into occount the soft photon emission vertex which is written in the
textbook for the scattering of charged particle by external electromagnetic
fied or collision of charged particles.Based on low-energy theorem the most
singular contribution for the matrix element $T_{n}=\left\langle \Omega
|\psi|r;k_{1},....,k_{n}\right\rangle $ is known as the soft photons attached
to external line.Ward-Takahashi-identity
\begin{align}
T_{n}  &  =\epsilon_{\mu}^{n}T_{n}^{\mu},\nonumber\\
k_{\mu}^{n}T_{n}^{\mu}(r;k_{1},..kn)  &  =eT_{n-1}(r;k_{1,}..k_{n-1}),k_{\mu
}^{n2}\neq0
\end{align}
is proved to be satisfied with the use of LSZ reduction formula[2].We have an
approximate solution of (4)
\begin{equation}
T_{n}|_{k_{n}^{2}=0}=e\frac{\gamma\cdot\epsilon}{\gamma\cdot(r+k_{n}%
)-m}T_{n-1}.
\end{equation}
From this relation the n-photon matrix element is replaced by the products of
$T_{1}$
\begin{equation}
T_{n}T_{n}^{+}\gamma_{0}\rightarrow%
{\displaystyle\prod\limits_{j=1}^{n}}
T_{1}(k_{j})T_{1}^{+}(k_{j})\gamma_{0}.
\end{equation}
By one-photon matrix element[4]
\begin{equation}
T_{1}=-ie\frac{\gamma\cdot(r+k)+m}{(r+k)^{2}-m^{2}+i\epsilon}\gamma_{\mu
}\epsilon_{\lambda}^{\mu}(k)U_{S}(r),
\end{equation}
and the function
\begin{equation}
F=\int\frac{d^{3}k}{(2\pi)^{2}}\exp(ik\cdot x)\theta(k_{0})\delta(k^{2}%
)\sum\limits_{\lambda,S}T_{1}\overline{T_{1}},
\end{equation}
where polarization sum we have%
\begin{equation}
\Pi^{\mu\nu}(k)=\sum_{\lambda}\epsilon_{\lambda}^{\mu}(k)\epsilon_{\lambda
}^{\nu}(k)=-(g^{\mu\nu}-\frac{k^{\mu}k^{\nu}}{k^{2}})-d\frac{k^{\mu}k^{\nu}%
}{k^{2}}.
\end{equation}
The infinite sum $\sum_{n=0}T_{n}T_{n}^{+}\gamma_{0}/n!$ leads an $\exp(F)$.In
this way we obtain a dressed fermion propagator with soft photon
\begin{equation}
S_{F}(x)=-(\frac{i\gamma\cdot\partial}{m}+1)\int\frac{md^{2}r}{(2\pi)^{2}%
\sqrt{r^{2}+m^{2}}}\exp(ir\cdot x)\exp(F),
\end{equation}
where $F$ is known as model independent
\begin{equation}
F=-\frac{e^{2}}{2}(\frac{\gamma\cdot r+m}{m})\int\frac{d^{3}k}{(2\pi)^{2}}%
\exp(ik\cdot x)\theta(k^{0})\delta(k^{2})[\frac{m^{2}}{(r\cdot k)^{2}}%
+\frac{1}{(r\cdot k)}+\frac{d-1}{k^{2}}],
\end{equation}
here we used covaiant$\ d$ gauge photon propagator and $\delta(k^{2})$ is read
as the imaginary part of the free photon propagator at on shell.In our
approximation two kinds of spectral function satisfy $\rho_{2}=m\rho_{1}.$

\subsection{\bigskip Photon}

For unquenched case we use the dressed photon with massive fermion loop with
$N$ \ flavours.Spectral functions for dressed photon are given by vacuum
polarization [5,6]%
\begin{align}
\Pi_{\mu\nu}(k) &  \equiv ie^{2}\int\overline{d^{3}}pTr(\gamma_{\mu}\frac
{1}{\gamma\cdot p-m}\gamma_{\nu}\frac{1}{\gamma\cdot(p-k)-m})\nonumber\\
&  =-e^{2}\frac{T_{\mu\nu}}{8\pi}[(\sqrt{k^{2}}+\frac{4m^{2}}{\sqrt{k^{2}}%
})\ln(\frac{2m+\sqrt{k^{2}}}{2m-\sqrt{k^{2}}})-4m],\\
T_{\mu\nu} &  =(g_{\mu\nu}-\frac{k_{\mu}k_{\nu}}{k^{2}}),\overline{d^{3}%
}p=\frac{d^{3}p}{(2\pi)^{3}},\nonumber
\end{align}%
\begin{equation}
D_{\mu\nu}^{-1}(k)\equiv-(T_{\mu\nu}k^{2}+dk_{\mu}k_{\nu})+\Pi_{\mu\nu}(k).
\end{equation}
Polarization function $\Pi(k)$ is
\begin{align}
\Pi(k) &  =-\frac{e^{2}}{8\pi}[(\sqrt{-k^{2}}+\frac{4m^{2}}{\sqrt{-k^{2}}}%
)\ln(\frac{2m+\sqrt{-k^{2}}}{2m-\sqrt{-k^{2}}})-4m],\nonumber\\
&  =-\frac{e^{2}}{8}i\sqrt{-k^{2}}(-k^{2}>0,m=0),\\
&  =\frac{e^{2}}{6\pi m}k^{2}+O(k^{4})(-k^{2}/m\ll1).
\end{align}
Fermion mass is assumed to be generated dynamically.In quenched case it is
shown that $m$ is proportional to $e^{2}$[2].For massless case or high-energy
limit we have for number of $N$ fermion flavour
\begin{equation}
\rho_{\gamma}^{D}(k)=\frac{1}{\pi}\operatorname{Im}D_{F}(k)=\frac{c\sqrt
{k^{2}}}{k^{2}(k^{2}+c^{2})},c=\frac{e^{2}N}{8},
\end{equation}%
\begin{equation}
D_{\mu\nu}(k)=-T_{\mu\nu}\int_{0}^{\infty}\frac{\rho_{\gamma}(\mu^{2})d\mu
^{2}}{k^{2}-\mu^{2}+i\epsilon}-d\frac{k_{\mu}k_{\nu}}{(k^{2}-i\epsilon)^{2}}.
\end{equation}
If we include finite threshold effects of massive fermion pair we have
\begin{equation}
\rho_{\gamma}^{F}(k)=\frac{1}{\pi}\Im D_{F}(k)=\delta(k^{2})+\frac{1}{\pi}%
\Im\frac{1}{(-k^{2}-\Pi(-k^{2}))}\theta(-k^{2}-4m^{2}).
\end{equation}

\section{Analysis in position space}

To evaluate the function $F$ it is helpful to use the exponential
cut-off(infrared cut-off)[2,3].Using the bare photon propagator with bare mass
$\mu$ ;$D_{F}^{(0)}(x)_{+}=\exp(-\mu|x|)/8\pi i|x|.$The function $F$ is
written in the following form%
\begin{equation}
F=ie^{2}m^{2}\int_{0}^{\infty}\alpha d\alpha D_{F}(x+\alpha r)-e^{2}\int
_{0}^{\infty}d\alpha D_{F}(x+\alpha r)-i(d-1)e^{2}\frac{\partial}{\partial
\mu^{2}}D_{F}(x,\mu^{2}).
\end{equation}
The above formulea are derived by the parameter tric%
\begin{align}
\lim_{\epsilon\rightarrow0}\int_{0}^{\infty}d\alpha\exp(-\alpha(\epsilon
-ik\cdot r))  &  =\frac{i}{k\cdot r},\nonumber\\
\lim_{\epsilon\rightarrow0}\int_{0}^{\infty}\alpha d\alpha\exp(-\alpha
(\epsilon-ik\cdot r))  &  =-\frac{1}{(k\cdot r)^{2}}.
\end{align}
In this case we have%
\begin{align}
F  &  =\frac{e^{2}}{8\pi}[\frac{\exp(-\mu|x|)-\mu|x|E_{1}(\mu|x|)}{\mu}%
-\frac{E_{1}(\mu|x|)}{m}\nonumber\\
&  +\frac{(d-1)\exp(-\mu\left\vert x\right\vert ))}{2\mu}],|x|=\sqrt{x^{2}}%
\end{align}
where $\mu$ is a bare photon mass.Short distance behaiviour of $F$ has the
following form%
\begin{equation}
F\sim\frac{e^{2}(1+d)}{16\pi\mu}+\frac{e^{2}}{8\pi m}(\gamma+(1+m|x|)\ln
(\mu\left\vert x\right\vert ))-\frac{(d+1-2\gamma)e^{2}|x|}{16\pi},(\mu
|x|\ll1).
\end{equation}
Long distance behaviour is given by the asymptotic expansion of $E_{1}%
(\mu|x|)$%
\begin{equation}
E_{1}(z)\sim\frac{\exp(-z)}{z}\{1-\frac{1}{z}+\frac{1\cdot2}{z^{2}}%
-\frac{1\cdot2\cdot3}{z^{3}}+...\},(|\arg z|<\frac{3}{2}\pi),
\end{equation}%
\begin{equation}
F\sim-\frac{e^{2}}{8\pi}[\frac{\exp(-\mu|x|)}{\mu^{2}|x|}+\frac{\exp(-\mu
|x|)}{m\mu|x|}+\frac{(d-1)\exp(-\mu|x|)}{2\mu}],(\mu|x|\gg1).
\end{equation}
where $\gamma$ is an Euler constant.In (20) linear term in $\left\vert
x\right\vert $ is understood as the finite mass shift from the form of the
propagator in position space (9,25) and $\left\vert x\right\vert \ln
(\mu\left\vert x\right\vert )$ term is position dependent mass
\begin{align}
m  &  =m_{0}+\frac{e^{2}}{16\pi}(1+d-2\gamma)+\frac{e^{2}}{8\pi m}\mu,\\
m(x)  &  =m-\frac{e^{2}}{8\pi}\ln(\mu\left\vert x\right\vert ),
\end{align}
which has mass changing effects at short distance and it will be discussed in
section 4.These mass terms has different gauge dependence from that obtained
by self-energy in the Dyson-Schwinger equation.Here we notice that the
propagator in poition space can be written%
\begin{align}
S_{F}(x)  &  =-(i\gamma\cdot\partial+m)\frac{1}{4\pi|x|}\left(
\begin{array}
[c]{c}%
\exp(-m|x|)A(\mu|x|)^{D+C|x|}(\mu|x|\ll1)\\
\exp(-m|x|)(\mu|x|\gg1)
\end{array}
\right)  ,\nonumber\\
&  =i\gamma\cdot\partial S_{V}(x)+S_{S}(x),
\end{align}
where%
\begin{equation}
A=\exp(\frac{e^{2}(1+d)}{16\pi\mu}+\frac{e^{2}\gamma}{8\pi m}),C=\frac{e^{2}%
}{8\pi},D=\frac{e^{2}}{8\pi m}.
\end{equation}
For the finitenes of the value $S_{F}(0)$ we imply $D=1.$In this case the
physical mass equals to $m=e^{2}/8\pi$ and fermion may be confined for finite
$\mu$.Here we apply the spectral function of photon to evaluate the unquenched
fermion proagator.We simply integrate the function $F(x,\mu)$ for quenched
case which is given in (19),where $\mu$ is a photon mass.Spectral function of
photon with massless fermion loop is given in (13) and we have
\begin{equation}
\rho_{\gamma}^{D}(\mu)=\frac{c}{\pi\mu(\mu^{2}+c^{2})},Z_{3}^{-1}=\int
_{0}^{\infty}2\rho_{\gamma}^{D}(\mu)\mu d\mu=1.
\end{equation}
In this case the spectral function of the fermion is given
\begin{equation}
\widetilde{\rho}(x)=\int_{0}^{\infty}2\rho_{\gamma}^{D}(\mu)\exp(F(x,\mu
))d\mu.
\end{equation}
In this way the short distance fermion propagator with $N$ flavours is
modified in the gauge $d=-1$
\begin{equation}
S_{F}(x)=-(i\gamma\cdot\partial+m)\frac{\exp(-m\left\vert x\right\vert )}%
{4\pi\left\vert x\right\vert }\widetilde{\rho}(x).
\end{equation}
For the case of $N$ fermion flavour we assume the physical mass $m$ as
$m=c/N\pi.$In the whole region of $|x|$ we evaluate $\widetilde{\rho}(x)$
numerically including finite threshold effect for the photon spectral function
with massive fermion loop
\begin{equation}
\widetilde{\rho}(x)=\int_{4m^{2}}^{\infty}\exp(F(s,x))\rho_{\gamma}^{F}(s)ds.
\end{equation}
The renormalization constant $Z_{3}^{-1}$ is defined as the residue of pole
\begin{equation}
Z_{3}^{-1}=1+\int_{4m^{2}}^{\infty}\rho_{\gamma}^{F}(s)ds.
\end{equation}
We set the mass $m=c/N\pi$ and see the $N$ dependence of $Z_{3}^{-1}$ for weak
coupling $c=1/8$ in Fig.1.In Fig.2 we see the screenig effect leads infrared
finite spectral function with dynamical fermion mass $m=c/N\pi$ from $N=1$ to
$4$ in the gauge $d=-1,c=1/8.$ In comparison with quenched case with finite
cut-off $\mu$ these are reduced from unity by screening effects at large distance.

.%
{\parbox[b]{3.0113in}{\begin{center}
\includegraphics[
trim=0.000000in 0.000000in 0.347671in 0.347671in,
height=3.0113in,
width=3.0113in
]%
{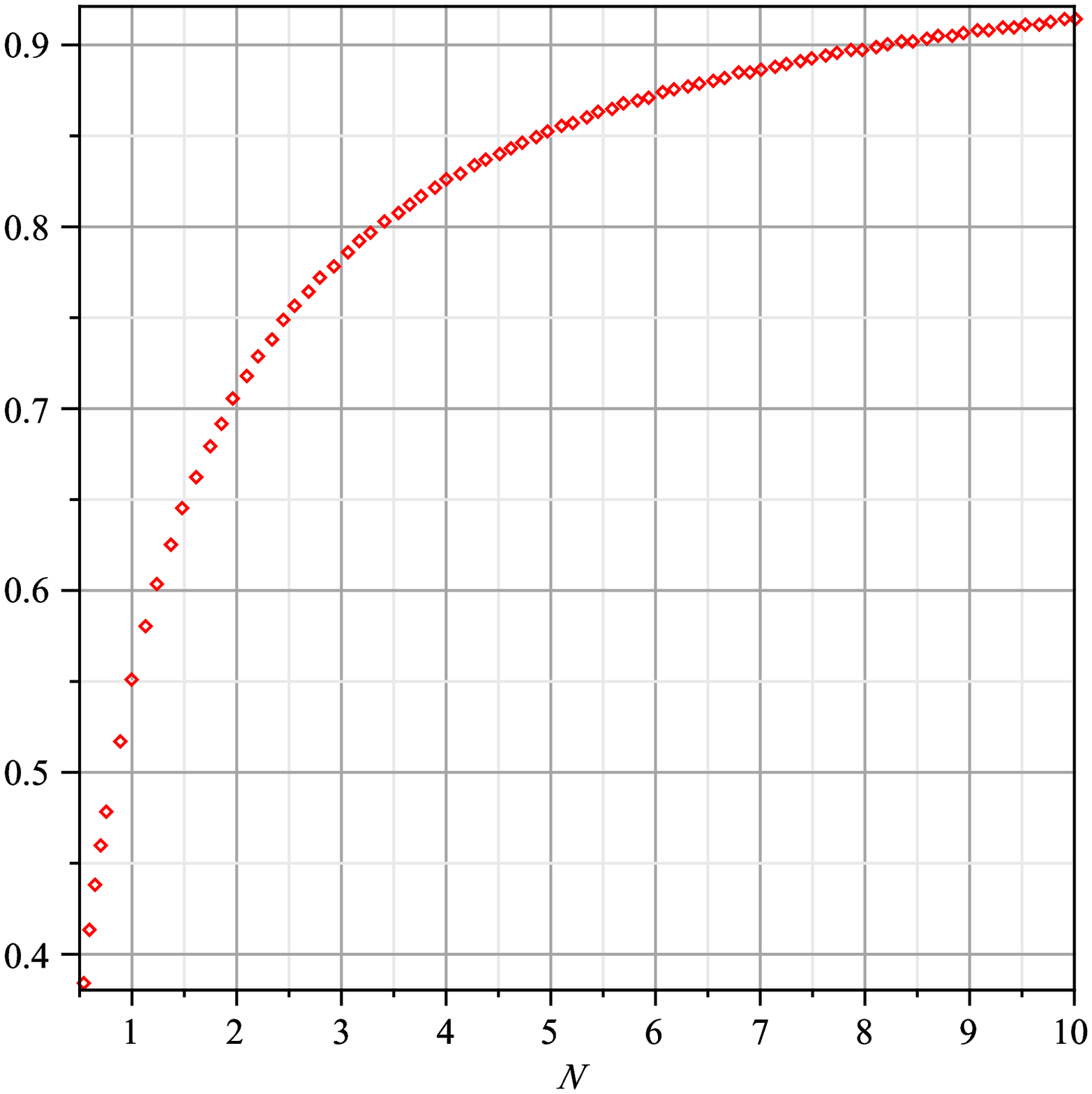}%
\\
Fig.1 $Z_{3}^{-1}-1$ for $N=1/2..10,c=1/8.$%
\end{center}}}%
{\parbox[b]{3.0113in}{\begin{center}
\includegraphics[
height=3.0113in,
width=3.0113in
]%
{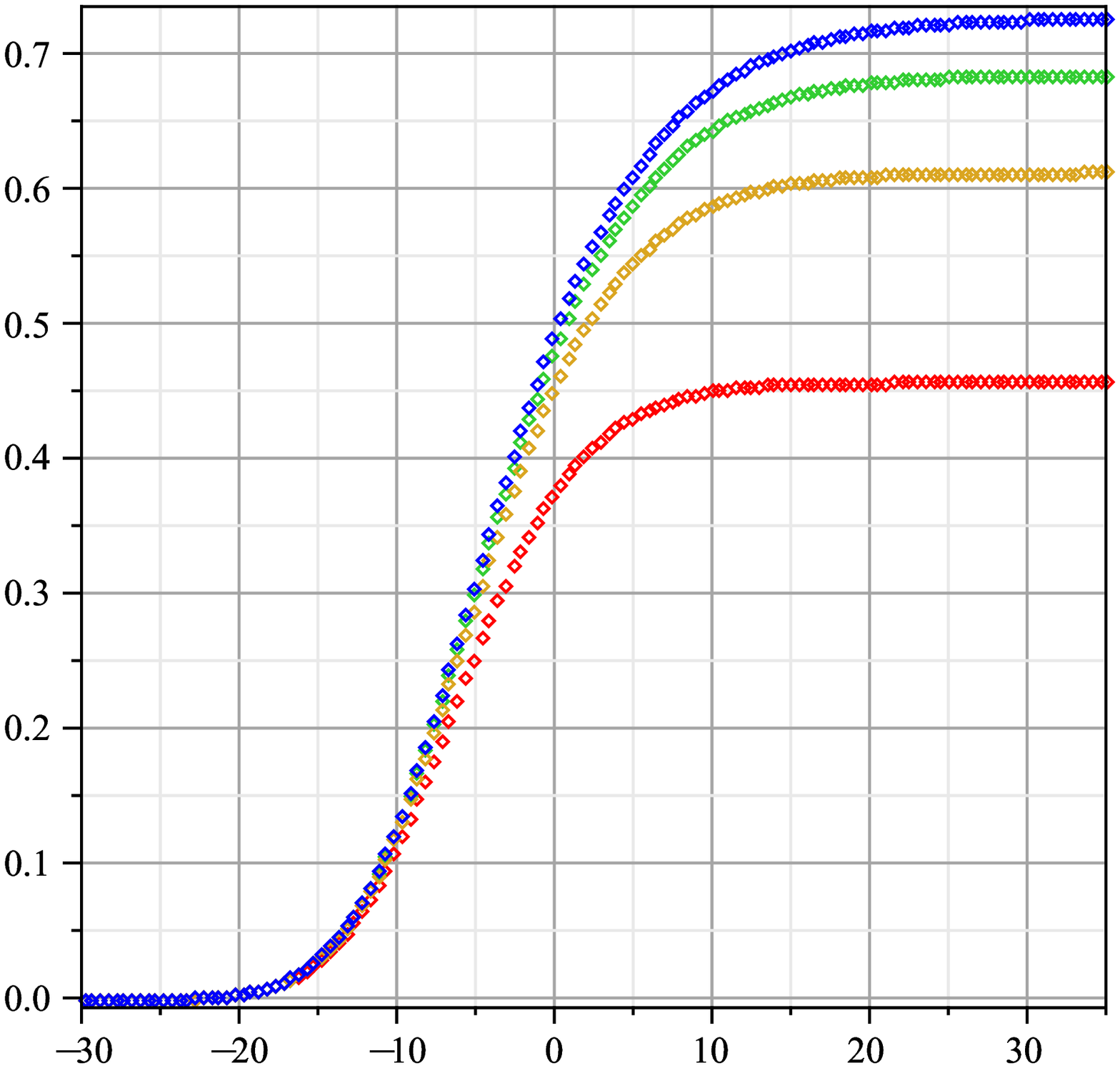}%
\\
Fig.2.$\widetilde{\rho}(x)$ for $c=1/8,N=1(bottom),2,3,4(top)|.$%
\end{center}}}%

\section{Analysises in momentum space}

\subsection{structure in Euclid space}

Now we turn to the fermion propagator in momentum space.The momentum space
propagator at short distance is given by Fourier transform%
\begin{equation}
S_{F}(p)=-\int d^{3}x\exp(-ip\cdot x)(i\gamma\cdot\partial+m)\frac
{\exp(-m\left\vert x\right\vert }{4\pi|x|}\exp(F(x)),
\end{equation}
where
\begin{equation}
\exp(F(x))=A\exp(D\gamma)(\mu\left\vert x\right\vert )^{D+C\left\vert
x\right\vert },D=\frac{c}{N\pi m},C=\frac{c}{N}.
\end{equation}
First we examine the perturbative effects by expanding $F$ in powers of
$e^{2}$ for quenched case at short distace%
\begin{align}
\exp(F)  &  =1+F(x)=1+A-B|x|+C|x|\ln(\mu|x|)+D\ln(\mu|x|),\nonumber\\
A  &  =\exp(\frac{e^{2}(1+d)}{16\pi\mu}+\frac{e^{2}\gamma}{8\pi m}%
),B=\frac{(d+1-2\gamma)}{16\pi},C=\frac{e^{2}}{8\pi},D=\frac{e^{2}}{8\pi m}.
\end{align}
Terms proportional to $B,C$ represents dynamical mass generation correctly.On
the other hand it is known that%
\begin{equation}
\int d^{3}x\exp(-ip\cdot x)\frac{\exp(-m\left\vert x\right\vert )}%
{4\pi\left\vert x\right\vert }(\mu\left\vert x\right\vert )^{D}=\frac{2m\mu
}{(p^{2}+m^{2})^{2}}\text{ for }D=1.\nonumber
\end{equation}
for Euclidean momentum $p^{2}\geq0.$However above formule are restricted for
small $|x|,$we may include the finite range effect and test the high ernergy
behaviour for small $\mu$
\begin{align}
&  \int_{0}^{1/\mu}\frac{\exp(-m|x|)}{|x|}(\mu|x|)\frac{\sin(p|x|)}{p|x|}%
x^{2}dx\nonumber\\
&  \rightarrow-\frac{\mu\exp(-m/\mu)\cos(p/\mu)}{p^{2}}+\frac{\mu\exp
(-m/\mu)\sin(p/\mu)(m-\mu)}{p^{3}}+O(1/p^{4}).
\end{align}
Above formula shows the oscillation and the propagator does not dump as
$1/p^{4}$.For large $N$ this oscillation effects becomes small as
$\mu\rightarrow0.$Numerical solutions of the scalar part of the propagator
$S_{S}(p)$ with various $N$ are shown in Fig.3-1, Fig.3-2, for
$c=1/8,1,N=1,2,3$ respectively.The boundary condition at $p=0$ is not specfied
in our approximation.For weak coupling $c=1/8$ and $c=1$ zero momentum mass
for $N=1$ is the largest among them, which has been seen in the
Dyson-Schwinger analysis[8].

\bigskip%
{\parbox[b]{2.9862in}{\begin{center}
\includegraphics[
height=2.9862in,
width=2.9862in
]%
{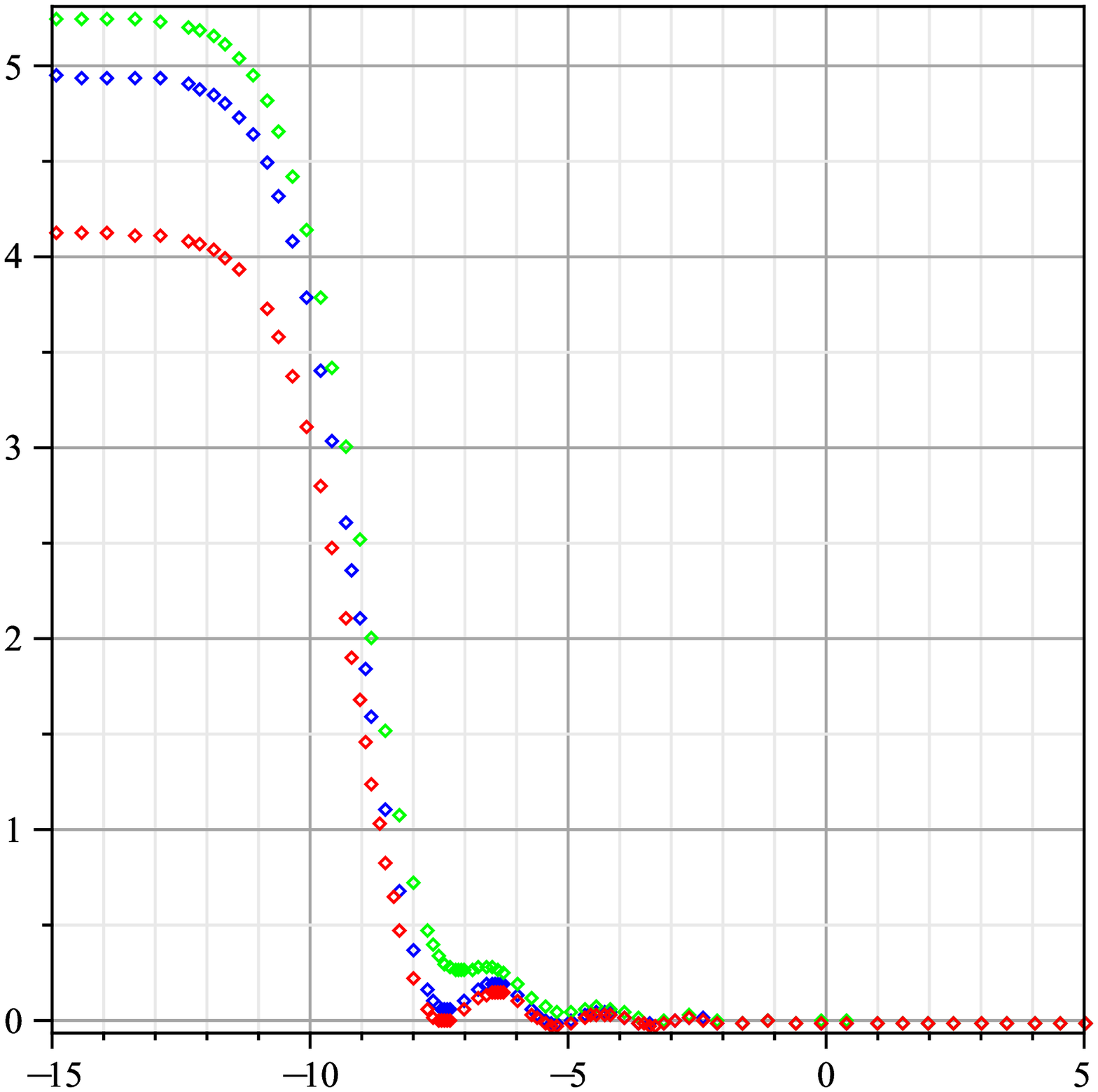}%
\\
Figure.3-1 $S_{S}(p)$ for $c=1/8,$ $N=1(top),2(middle),3(bottom),p=\exp
(\pi/2\sinh(k/5)).$%
\end{center}}}%
{\parbox[b]{3.0113in}{\begin{center}
\includegraphics[
height=3.0113in,
width=3.0113in
]%
{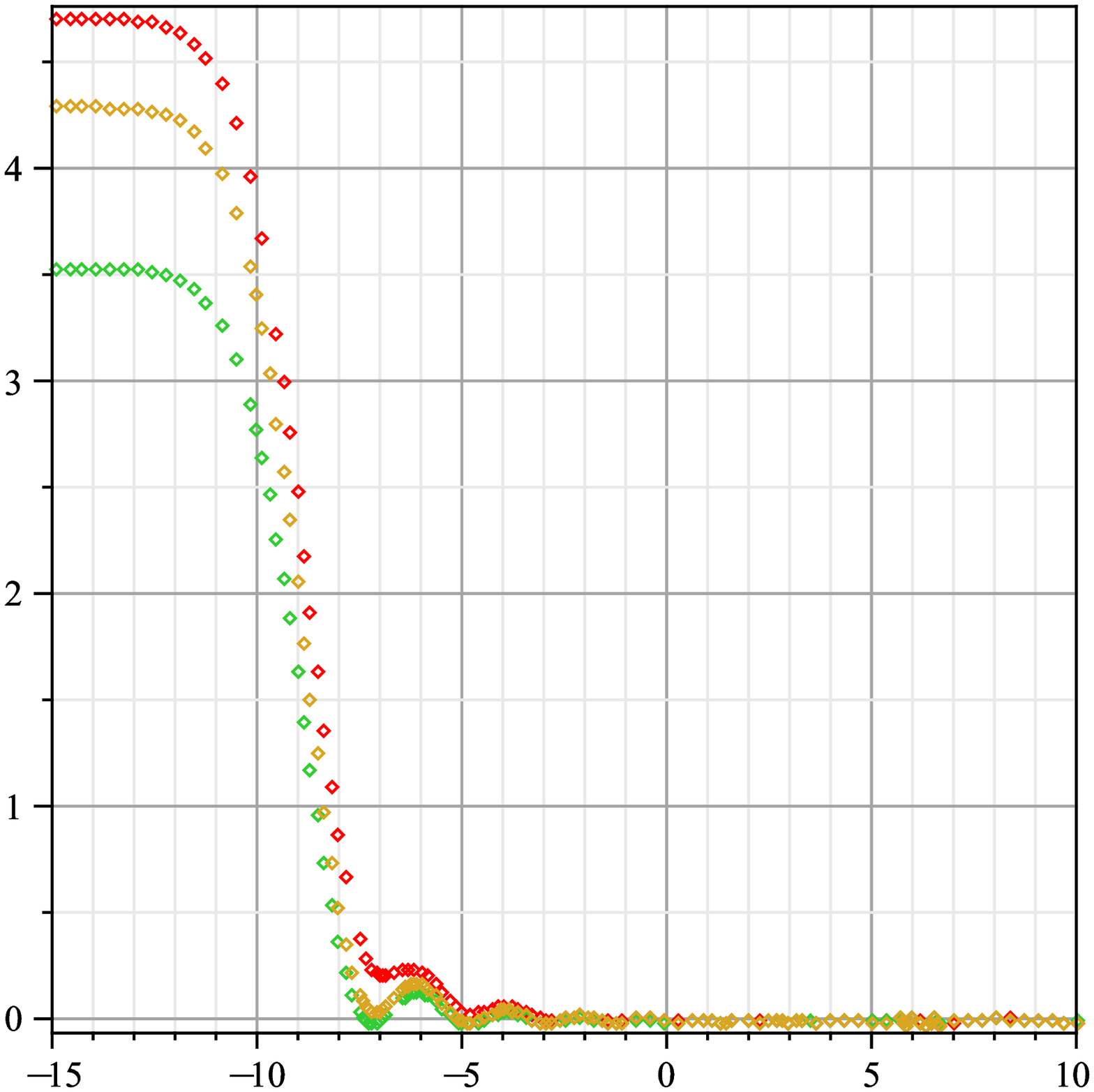}%
\\
Figure.3-2 $S_{S}(p)$ for $c=1,$ $N=1(bottom),2(middle),3(top),p=\exp
(\pi/2\sinh((m/5))$%
\end{center}}}%

\subsection{\bigskip structure in Minkowski space}

Now we derive the propagator in Minkowski momentum region.To do this it is
helpful in real time to derive the spectral function[1,9].In Minkowski space
we change the variable from $r=\sqrt{x^{2}}$ to $iT.$We may define
\begin{equation}
\rho(s^{\prime2})=\frac{1}{2\pi}\int_{-\infty}^{\infty}\exp(-i(s^{\prime
2}/m^{2}-1)T)\operatorname{Im}(\exp(\widetilde{F}(iT))dT,
\end{equation}
which is normalized to $\delta(s^{2}/m^{2}-1)$ for free case where
\begin{equation}
\exp(\widetilde{F}(iT))=\int_{4m^{2}}^{\infty}d\sigma\rho^{F}(\sigma
)\exp(F(iT,\sqrt{\sigma})).
\end{equation}
It is understood that the Fourier transformation is performed in $s^{2}$ by
the following form
\begin{equation}
\frac{1}{2\pi}\int_{-\infty}^{\infty}dT\exp(-i(m-s)T)+\exp
(-i(m+s)T)=2|s|\delta(s^{2}-m^{2}).
\end{equation}
Taking the imaginary part of the function $\exp(F(\mu,iT))$ at short distance
in the gauge $d=-1$ for quenched case for $D=1$
\begin{equation}
\operatorname{Im}\exp(F(\mu,iT))\approx\exp(-\frac{\pi}{2}CT)\mu T\cos
(CT\ln(\mu T)),
\end{equation}
provided
\begin{equation}
(i\mu T)^{iCT+D}=\exp(\frac{\pi}{2}(-CT+Di))(\mu T)^{D}(\cos(CT\ln(\mu
T))+i\sin(CT\ln(\mu T))).
\end{equation}
In this way we see the oscillation of the propagator in Minkowski space by the
effects of position dependent mass as $m=iCT\ln(iCT).$In Fig.4 we see the
profile of real and imaginary part of the function $\exp(F(iT))$ for $c=1,N=1$.

\ \ \
{\parbox[b]{3.0113in}{\begin{center}
\includegraphics[
height=3.0113in,
width=3.0113in
]%
{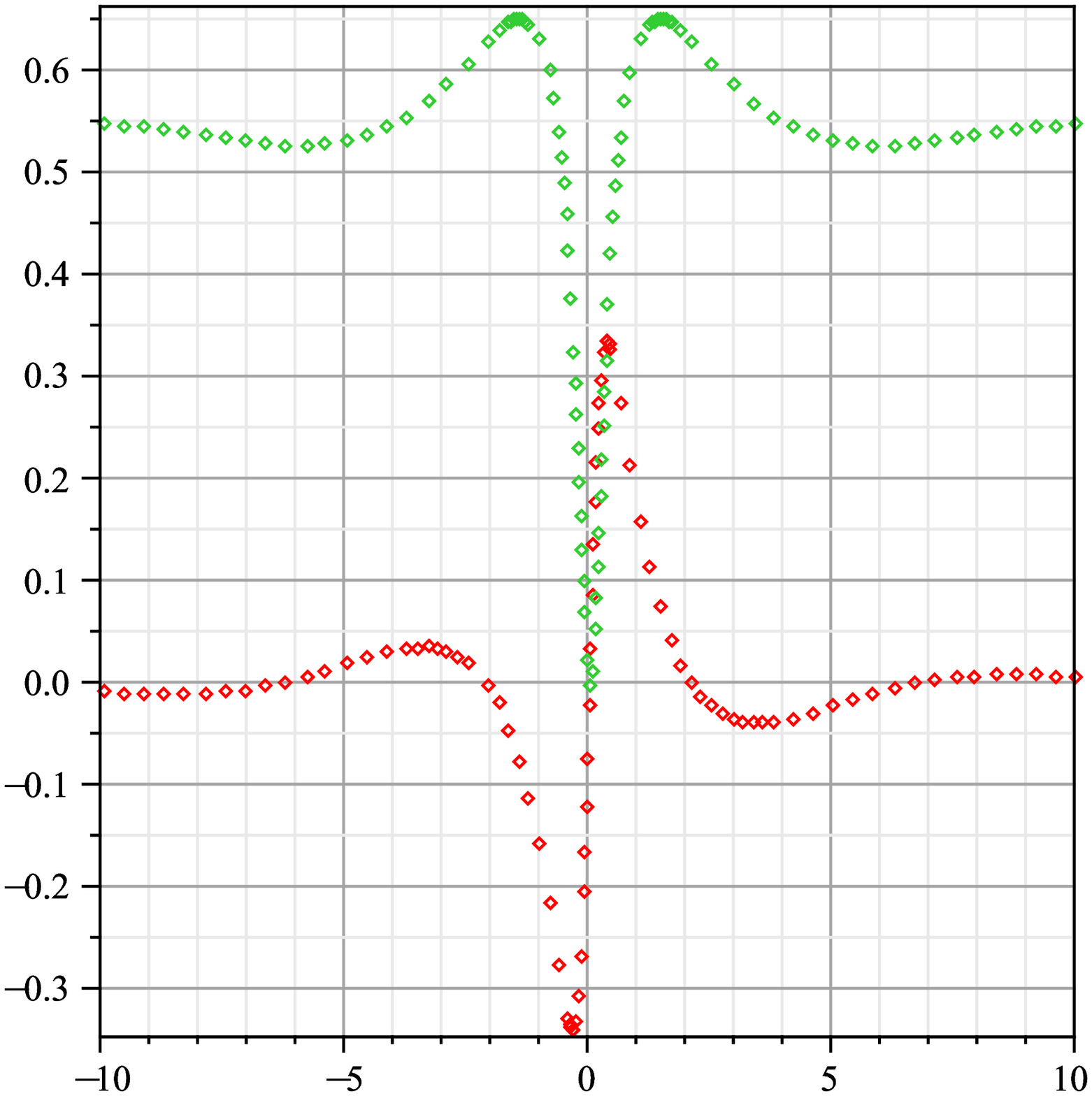}%
\\
Fig.4 $\Re(\exp(\widetilde{F}(iT))$, $\Im(\exp(\widetilde{F}(iT))..$%
\end{center}}}%
{\parbox[b]{3.0113in}{\begin{center}
\includegraphics[
trim=0.000000in 0.000000in 0.349213in 0.349213in,
height=3.0113in,
width=3.0113in
]%
{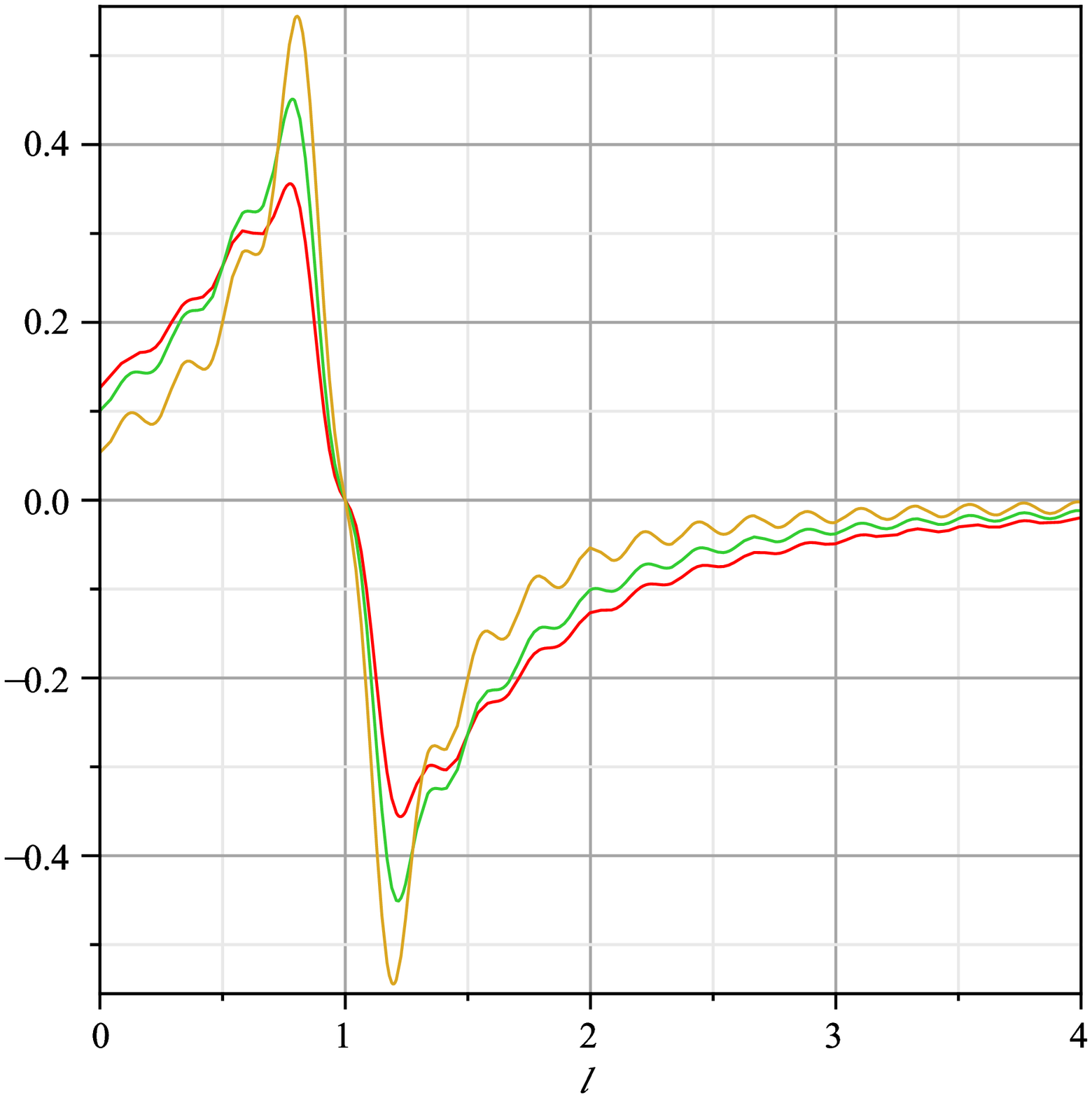}%
\\
Fig.5 $\rho(s),N=1(bottom),2,3(top).$%
\end{center}}}%

In Fig.5 the spectral function $\rho(s)$ in unit of $e^{2}$ looks symmetric
about the point $s=1$.The function $\exp(F(iT))$ is approximated as $i\mu T$
\ for small $T$.This fact suggests the singularity of $\rho(s)$ such as the
$-\delta^{\prime}(s-1)$.

\section{Renormalization constant and order parameter}

In this section we consider the renormalization constant in our model.It is
easy to evaluate the renormalization constant and bare mass which are defined
by assuming multiplicative renormalization
\begin{equation}
\frac{Z_{2}^{-1}}{\gamma\cdot p-m_{0}}=S_{F}(p).
\end{equation}%
\begin{align}
Z_{2}^{-1}  &  =\lim_{p\rightarrow\infty}\frac{1}{4}tr(\gamma\cdot
pS_{F}(p))\nonumber\\
&  =\lim_{p\rightarrow\infty}\int\frac{p^{2}\rho(s)ds}{p^{2}-s+i\epsilon}%
=\int\rho(s)ds,
\end{align}
where $Z_{2\text{ }}$is defined for one particle state in the intial and final
state in a weak sense $\psi(x)_{t\rightarrow+\infty,-\infty}\rightarrow
\sqrt{Z_{2}}\psi(x)_{out,in}$
\begin{equation}
Z_{2}(2\pi)^{-2}\theta(p_{0})\delta(p^{2}-m^{2})\equiv|\left\langle
p|\psi(0)|0\right\rangle |_{p^{2}=m^{2}}^{2}.
\end{equation}
However pole part is absent in our approximation and $Z_{2\text{ }}$vanishes
for $D>0$ case.For $D=1$ we have%
\begin{align}
Z_{2}^{-1}  &  =\int_{0}^{\infty}ds\int_{-\infty}^{\infty}\frac{dt}{2\pi}%
\exp(-i(s-1)t)\Im(\exp(\widetilde{F}(it))\nonumber\\
&  =\int_{-\infty}^{\infty}dt\delta_{+}(t)\Im(\exp(\widetilde{F}(it))=0,
\end{align}
provided $\Im(\exp(\widetilde{F}(it))=0$ in the limit $t_{+}\rightarrow0$ by
eq(38)$.$Order parameter for each flavour $\left\langle \overline{\psi}%
\psi\right\rangle $ is given
\begin{equation}
\left\langle \overline{\psi}\psi\right\rangle =-trS_{F}(x).
\end{equation}
The value of $\left\langle \overline{\psi}\psi\right\rangle $ is
$3.1(1.5.,1.0)\times10^{-3}e^{4}$ for $N=1(2,3)$ in $1/N$ which may be
compared with $1.2(0.13,0.0002)\times10^{-3}e^{4}$ for $N=1(2,3)$ in the CP
vertex with massless loop correction [8].In our approximation $N$ dependence
is mild.This may be understood as the large screening effect at small
$N.$Recently we solved the Bethe-Salpeter equation in the ladder approximation
for axial-scalar which corresponds the psedoscalar in four dimension.For
massless boundstate solution we find the $D=e^{2}/4\pi m=1$ from the
normalization condition at short distance[9].\qquad

\section{Summary}

We evaluated the fermion propagator in three dimensional QED with dressed
photon by the dispersion method.In the evaluation of lowest order matrix
element for fermion spectral function we obtain finite mass shift,wave
function renormalization and gauge invariant position dependent mass with bare
photon mass $\mu$ as an infrared cut-off..To remove linear infrared
divergences we include finite threshold effects of massive fermion loop to the
photon spectral function.We set the coupling constant mass ratio $c/N\pi m$ to
be unity which is consistent with perurabative analysis of the mass at high
energy which is proportional to$1/p^{4}$.In this case the order parameter
$\left\langle \overline{\psi}\psi\right\rangle $ becomes finite.In our
approximation vacumm expectation value and infrared mass are not sensitive to
flavour number $N$ which may due to the screening effects of $Z_{3}^{-1\text{
}}$for small $N$.In Minkowski space$.$the spectral fuction is not positive
definite and has a form such as $-\delta^{\prime}(p^{2}/m^{2}-1)$.In the
strong coupling there seems to be no particle content in the model.In our
analysis chiral symmetry breaks dyamically by anomalous dimension.Our results
suggests the importance of anomalous dimension to understand confinement.

\section{Aknowledgement}

The author would like to than Professor Rober Delbourgo for critiques of
linear divegrgence.

\section{References}

\noindent\lbrack1]R.Jackiw,L.Soloviev,Phys.Rev.\textbf{137}%
.3(1968)1485.\newline[2]Y.Hoshino,\textbf{JHEP0409}:048(2004),whereIhave a
wrong sign of $|x|\ln(\mu|x|)$ term.\nolinebreak\newline%
[3]Y.Hoshino,Nucl.Phys.A790(2007)613c-618c.\newline%
[4]K.Nishijima,\textbf{Fields and Particles},W.A.BENJAMIN,INC(1969).\newline%
[5]A.B.Waites,R.Delbourgo,Int.J.Mod.Phys.\textbf{A7}(1992)6857.\newline%
[6]C.Itzykson,J.B.Zuber,Quantum field theory,McGRAW-HILL.\newline%
[7]J.Schwinger,\textbf{Particle Sources and Fields},vol.I,Pereseus Books
Publishing, L,L,C(1970).\newline%
[8]C.S.Fischer,R.Alkofer,T.Dahm,P.Maris,Phys.Rev.\textbf{D70}%
,073007(2004):[arXiv:hep-th/0407014].\newline[9]Y.Hoshino:[arXiv:0706.1063]

\end{document}